\documentclass[
    ,final            
  ,numberedheadings 
  ]
  {aipproc}

\layoutstyle{8x11single}

\def\eq#1{{Eq.~(\ref{#1})}}

\usepackage{latexsym}

\usepackage{graphicx,amssymb,amsmath}
\newcommand{\cc}{cosmological constant}
\newcommand{\del}{\partial}
\newcommand{\dphi}{\partial_i \phi \partial^i \phi}

\begin{document}


\title[Dark Energy]{ Dark Energy: Mystery of the Millennium }

                

\author{T.~Padmanabhan}{
  address={IUCAA, 
Post Bag 4, Ganeshkhind,\\
 Pune - 411 007, INDIA.\\
email: nabhan@iucaa.ernet.in}
}

\begin{abstract}

\noindent
Nearly seventy  per cent of the energy density in the universe is unclustered and exerts negative pressure.
This conclusion --- now supported by numerous observations --- poses the greatest challenge for theoretical physics today. I discuss this issue with  special emphasis on the cosmological constant as the possible choice for the dark energy. Several curious  features of a universe with a cosmological constant are
described and some possible approaches to understand the nature of the cosmological constant are reviewed. In particular, I show how some of the recent ideas, related to a thermodynamic route to gravity, allow us to: (i) create a paradigm in which the bulk value of cosmological constant is irrelevant and (ii) obtain the correct, observed, value for the cosmological constant from vacuum fluctuations in a region confined by the deSitter horizon. 
\end{abstract}

\maketitle


\section{ The Rise of the Dark Energy: Brief History}

The cosmological data of exquisite quality, which became available in the last couple of decades, have thrusted upon us a rather preposterous
composition for the universe which defies any simple explanation, thereby posing the greatest challenge
theoretical physics has ever faced.
It is conventional to measure
the energy densities of the various species which drive the expansion of the universe in terms of a \textit{critical energy density} $\rho_c=3H^2_0/8\pi G$  where $H_0=(\dot a/a)_0$
is the rate of expansion of the universe at present.  The variables $\Omega_i=\rho_i/\rho_c$ 
will then give the fractional contribution of different components of the universe ($i$ denoting baryons, dark matter, radiation, etc.) to the  critical density. Observations  suggest that the universe has
 $0.98\lesssim\Omega_{tot}\lesssim1.08$  with radiation (R), baryons (B), dark matter, made of weakly interacting massive particles (DM) and dark energy (DE) contributing  $\Omega_R\simeq 5\times 10^{-5},\Omega_B\simeq 0.04,\Omega_{DM}\simeq 0.26,\Omega_{DE}\simeq 0.7,$ respectively. All known observations \cite{cmbr,baryon,h}
are consistent with such an --- admittedly weird --- composition for the universe. 

Among all these components, the dark energy, which exerts negative pressure, is probably the weirdest. And nobody really wanted it! To understand its rapid acceptance by the community one needs to look at its recent history briefly.  Early analysis of several observations \cite{earlyde}
indicated that this component is unclustered and has negative pressure --- the observation which made me personally sit up and take note being the APM result. This is confirmed dramatically by the supernova observations\cite{sn}.   The current observations suggest that this component has 
$w=p/\rho\lesssim-0.78$
and contributes $\Omega_{DE}\cong 0.60-0.75$; for a critical look at the current data, see \cite{tptirthsn1}. 

While the composition of the universe is puzzling --- and I will concentrate on the \textit{unknown}, puzzling aspects of our universe for the rest of the talk --- it should not prevent us from appreciating  the remarkable \textit{successes} of the standard cosmological paradigm. 
The  key idea is that if there existed small fluctuations in the energy density in the early universe, then gravitational instability can amplify them in a well-understood manner  leading to structures like galaxies etc. today. The most popular model for generating these fluctuations is based on the idea that if the very early universe went through an inflationary phase \cite{inflation}, then the quantum fluctuations of the field driving the inflation can lead to energy density fluctuations\cite{genofpert,tplp}. It is possible to construct models of inflation such that these fluctuations are described by a Gaussian random field and are characterized by a power spectrum of the form $P(k)=A k^n$ with $n\simeq 1$. The models cannot predict the value of the amplitude $A$ in an unambiguous manner but it can be determined from CMBR observations. The CMBR observations are consistent with the inflationary model for the generation of perturbations and gives $A\simeq (28.3 h^{-1} Mpc)^4$ and $n=0.97\pm0.023$. (The first results were from COBE \cite{cobeanaly} and
WMAP has re-confirmed them with far greater accuracy).
When the perturbation is small, one can use well defined linear perturbation theory to study its growth
\cite{adcos}. But when $\delta\approx(\delta\rho/\rho)$ is comparable to unity the perturbation theory
breaks down. Since there is more power at small scales, smaller scales go non-linear first and structure forms hierarchically. 
The non linear evolution of the  \textit{dark matter halos} (which is an example of statistical mechanics
 of self gravitating systems; see e.g.\cite{smofgs}) can be understood by simulations as well as theoretical models based on approximate ansatz
\cite{nlapprox} and  nonlinear scaling relations \cite{nsr}.
 The baryons in the halo will cool and undergo collapse
 in a fairly complex manner because of gas dynamical processes. 
 It seems unlikely that the baryonic collapse and galaxy formation can be understood
 by analytic approximations; one needs to do high resolution computer simulations
 to make any progress \cite{baryonsimulations}. The results obtained from all these
 attempts are broadly consistent with observations. 
 So, to the zeroth order, the universe is characterized by just seven numbers: $h\approx 0.7$ describing the current rate of expansion; $\Omega_{DE}\simeq 0.7,\Omega_{DM}\simeq 0.26,\Omega_B\simeq 0.04,\Omega_R\simeq 5\times 10^{-5}$ giving the composition of the universe; the amplitude $A\simeq (28.3 h^{-1} Mpc)^4$ and the index $n\simeq 1$ of the initial perturbations. Establishing this cosmological paradigm is a  remarkable progress by any sensible criterion.
 
The remaining challenge, of course,  is to make some sense out of these numbers from a more fundamental point of view.
It is rather frustrating that the only component of the universe which we understand theoretically is the radiation! While understanding the
baryonic and dark matter components [in particular the values of $\Omega_B$ and $\Omega_{DM}$] is by no means trivial, the issue of dark energy is lot more perplexing, thereby justifying the attention it has received recently.

The key observational feature of dark energy is that --- treated as a fluid with a stress tensor $T^a_b=$ dia     $(\rho, -p, -p,-p)$ 
--- it has an equation state $p=w\rho$ with $w \lesssim -0.8$ at the present epoch. 
The spatial part  ${\bf g}$  of the geodesic acceleration (which measures the 
  relative acceleration of two geodesics in the spacetime) satisfies an \textit{exact} equation
  in general relativity  given by:
  \begin{equation}
  \nabla \cdot {\bf g} = - 4\pi G (\rho + 3p)
  \label{nextnine}
  \end{equation} 
 This  shows that the source of geodesic  acceleration is $(\rho + 3p)$ and not $\rho$.
  As long as $(\rho + 3p) > 0$, gravity remains attractive while $(\rho + 3p) <0$ can
  lead to repulsive gravitational effects. In other words, dark energy with sufficiently negative pressure will
  accelerate the expansion of the universe, once it starts dominating over the normal matter.  This is precisely what is established from the study of high redshift supernova, which can be used to determine the expansion
rate of the universe in the past \cite{sn}. 

The simplest model for  a fluid with negative pressure is the
cosmological constant (for a sample of recent reviews, see \cite{cc}) with $w=-1,\rho =-p=$ constant.
If the dark energy is indeed a cosmological constant, then it introduces a fundamental length scale in the theory $L_\Lambda\equiv H_\Lambda^{-1}$, related to the constant dark energy density $\rho_{_{\rm DE}}$ by 
$H_\Lambda^2\equiv (8\pi G\rho_{_{\rm DE}}/3)$.
In classical general relativity,
    based on the constants $G, c $ and $L_\Lambda$,  it
  is not possible to construct any dimensionless combination from these constants. But when one introduces the Planck constant, $\hbar$, it is  possible
  to form the dimensionless combination $H^2_\Lambda(G\hbar/c^3) \equiv  (L_P^2/L_\Lambda^2)$.
  Observations then require $(L_P^2/L_\Lambda^2) \lesssim 10^{-123}$.
  As has been mentioned several times in literature, this will require enormous fine tuning. What is more,
 in the past, the energy density of 
  normal matter and radiation  would have been higher while the energy density contributed by the  cosmological constant
  does not change.  Hence we need to adjust the energy densities
  of normal matter and cosmological constant in the early epoch very carefully so that
  $\rho_\Lambda\gtrsim \rho_{\rm NR}$ around the current epoch.
  This raises the second of the two cosmological constant problems:
  Why is $(\rho_\Lambda/ \rho_{\rm NR}) = \mathcal{O} (1)$ at the 
  {\it current} phase of the universe ?

\section{ Scalar Fields: The `Denial' approach to cosmological constant}    
  
  Because of these conceptual problems associated with the cosmological constant, people have explored a large variety of alternative possibilities. The most popular among them uses a scalar field $\phi$ with a suitably chosen potential $V(\phi)$ so as to make the vacuum energy vary with time. The hope then is that, one can find a model in which the current value can be explained naturally without any fine tuning.
  A simple form of the source with variable $w$ are   scalar fields with
  Lagrangians of different forms, of which we will discuss two possibilities:
    \begin{equation}
  L_{\rm quin} = \frac{1}{2} \partial_a \phi \partial^a \phi - V(\phi); \quad L_{\rm tach}
  = -V(\phi) [1-\partial_a\phi\partial^a\phi]^{1/2}
  \label{lquineq}
  \end{equation}
  Both these Lagrangians involve one arbitrary function $V(\phi)$. The first one,
  $L_{\rm quin}$,  which is a natural generalization of the Lagrangian for
  a non-relativistic particle, $L=(1/2)\dot q^2 -V(q)$, is usually called quintessence (for
  a small sample of models, see \cite{phiindustry}).
    When it acts as a source in Friedman universe,
   it is characterized by a time dependent
  $w(t)$ with
    \begin{equation}
  \rho_q(t) = \frac{1}{2} \dot\phi^2 + V; \quad p_q(t) = \frac{1}{2} \dot\phi^2 - V; \quad w_q
  = \frac{1-(2V/\dot\phi^2)}{1+ (2V/\dot\phi^2)}
  \label{quintdetail}
  \end{equation}

The structure of the second Lagrangian in Eq.~(\ref{lquineq}) (which arises in string theory \cite{asen})  can be understood by a simple analogy from
special relativity. A relativistic particle with  (one dimensional) position
$q(t)$ and mass $m$ is described by the Lagrangian $L = -m \sqrt{1-\dot q^2}$.
It has the energy $E = m/  \sqrt{1-\dot q^2}$ and momentum $k = m \dot
q/\sqrt{1-\dot q^2} $ which are related by $E^2 = k^2 + m^2$.  As is well
known, this allows the possibility of having \textit{massless} particles with finite
energy for which $E^2=k^2$. This is achieved by taking the limit of $m \to 0$
and $\dot q \to 1$, while keeping the ratio in $E = m/  \sqrt{1-\dot q^2}$
finite.  The momentum acquires a life of its own,  unconnected with the
velocity  $\dot q$, and the energy is expressed in terms of the  momentum
(rather than in terms of $\dot q$)  in the Hamiltonian formulation. We can now
construct a field theory by upgrading $q(t)$ to a field $\phi$. Relativistic
invariance now  requires $\phi $ to depend on both space and time [$\phi =
\phi(t, {\bf x})$] and $\dot q^2$ to be replaced by $\partial_i \phi \partial^i
\phi$. It is also possible now to treat the mass parameter $m$ as a function of
$\phi$, say, $V(\phi)$ thereby obtaining a field theoretic Lagrangian $L =-
V(\phi) \sqrt{1 - \del^i \phi \del_i \phi}$. The Hamiltonian  structure of this
theory is algebraically very similar to the special  relativistic example  we
started with. In particular, the theory allows solutions in which $V\to 0$,
$\dphi \to 1$ simultaneously, keeping the energy (density) finite.  Such
solutions will have finite momentum density (analogous to a massless particle
with finite  momentum $k$) and energy density. Since the solutions can now
depend on both space and time (unlike the special relativistic example in which
$q$ depended only on time), the momentum density can be an arbitrary function
of the spatial coordinate. The structure of this Lagrangian is similar to those analyzed in a wide class of models
   called {\it K-essence} \cite{kessence} and  provides a rich gamut of possibilities in the
context of cosmology
 \cite{tptachyon,tachyon}.

   Since  the quintessence field (or the tachyonic field)   has
   an undetermined free function $V(\phi)$, it is possible to choose this function
  in order to produce a given $H(a)$.
  To see this explicitly, let
   us assume that the universe has two forms of energy density with $\rho(a) =\rho_{\rm known}
  (a) + \rho_\phi(a)$ where $\rho_{\rm known}(a)$ arises from any known forms of source 
  (matter, radiation, ...) and
  $\rho_\phi(a) $ is due to a scalar field.  
  Let us first consider quintessence. Here,  the potential is given implicitly by the form
  \cite{ellis,tptachyon}
  \begin{equation}
  V(a) = \frac{1}{16\pi G} H (1-Q)\left[6H + 2aH' - \frac{aH Q'}{1-Q}\right]
  \label{voft}
   \end{equation} 
    \begin{equation}
    \phi (a) =  \left[ \frac{1}{8\pi G}\right]^{1/2} \int \frac{da}{a}
     \left[ aQ' - (1-Q)\frac{d \ln H^2}{d\ln a}\right]^{1/2}
    \label{phioft}
    \end{equation} 
   where $Q (a) \equiv [8\pi G \rho_{\rm known}(a) / 3H^2(a)]$ and prime denotes differentiation with respect to $a$.
   Given any
   $H(a),Q(a)$, these equations determine $V(a)$ and $\phi(a)$ and thus the potential $V(\phi)$. 
   \textit{Every quintessence model studied in the literature can be obtained from these equations.}
  
  Similar results exists for the tachyonic scalar field as well \cite{tptachyon}. For example, given
  any $H(a)$, one can construct a tachyonic potential $V(\phi)$ so that the scalar field is the 
  source for the cosmology. The equations determining $V(\phi)$  are now given by:
  \begin{equation}
  \phi(a) = \int \frac{da}{aH} \left(\frac{aQ'}{3(1-Q)}
   -{2\over 3}{a H'\over H}\right)^{1/2}
  \label{finalone}
  \end{equation}
   \begin{equation}
   V(a) = {3H^2 \over 8\pi G}(1-Q) \left( 1 + {2\over 3}{a H'\over H}-\frac{aQ'}{3(1-Q)}\right)^{1/2}
   \label{finaltwo}
   \end{equation}
   Equations (\ref{finalone}) and (\ref{finaltwo}) completely solve the problem. Given any
   $H(a)$, these equations determine $V(a)$ and $\phi(a)$ and thus the potential $V(\phi)$. 
A wide variety of phenomenological models with time dependent
  \cc\ have been considered in the literature; all of these can be 
   mapped to a 
  scalar field model with a suitable $V(\phi)$.

  While the scalar field models enjoy considerable popularity (one reason being they are easy to construct!)
  it is very doubtful whether they have helped us to understand the nature of the dark energy
  at any deeper level. These
  models, viewed objectively, suffer from several shortcomings:
   \begin{figure}[ht]
 \includegraphics[scale=0.5]{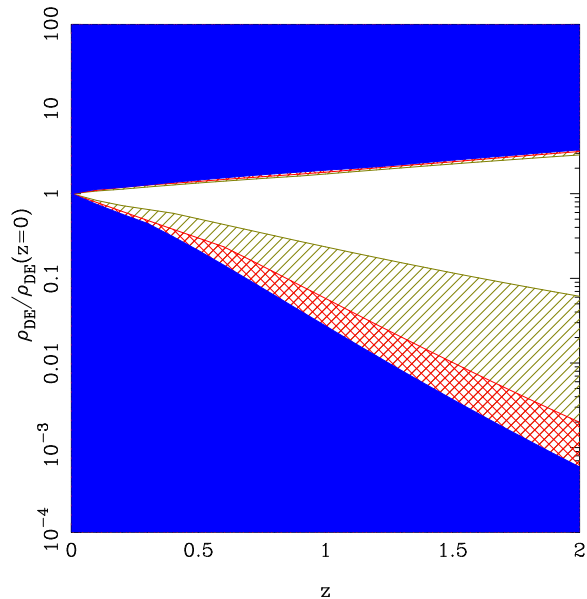}
 \caption{
 The observational constraints on the variation of dark energy density  as a function of
 redshift  from WMAP and SNLS data (see \cite{jbp}). The green/hatched region is
 excluded  at $68\%$ confidence limit, red/cross-hatched region at $95\%$
 confidence level  and the blue/solid region at $99\%$ confidence limit. The
 white region shows the allowed range of variation of dark energy at $68\%$
 confidence limit.  
  }
 \label{fig:bjp2ps}
 \end{figure} 
 \begin{itemize}
  \item
  They completely lack predictive power. As explicitly demonstrated above, virtually every form of $a(t)$ can be modeled by a suitable ``designer" $V(\phi)$.
  \item
  These models are  degenerate in another sense. The previous discussion  illustrates that even when $w(a)$ is known/specified, it is not possible to proceed further and determine
  the nature of the scalar field Lagrangian. The explicit examples given above show that there
  are {\em at least} two different forms of scalar field Lagrangians (corresponding to
  the quintessence or the tachyonic field) which could lead to
  the same $w(a)$. (See the first paper in ref.\cite{tptirthsn1} for an explicit example of such a construction.)
  \item
  All the scalar field potentials require fine tuning of the parameters in order to be viable. This is obvious in the quintessence models in which adding a constant to the potential is the same as invoking a \cc. So to make the quintessence models work, \textit{we first need to assume the \cc\ is zero.} These models, therefore, merely push the cosmological constant problem to another level, making it somebody else's problem!.
  \item
  By and large, the potentials  used in the literature have no natural field theoretical justification. All of them are non-renormalisable in the conventional sense and have to be interpreted as a low energy effective potential in an ad hoc manner.
  \item
  One key difference between \cc\ and scalar field models is that the latter lead to a $w(a)$ which varies with time. If observations have demanded this, or even if observations have ruled out $w=-1$ at the present epoch,
  then one would have been forced to take alternative models seriously. However, all available observations are consistent with \cc\ ($w=-1$) and --- in fact --- the possible variation of $w$ is strongly constrained \cite{jbp} as shown in Figure \ref{fig:bjp2ps}. 
  \item
  While on the topic of observational constraints on $w(t)$, it must be stressed that: (a) There is fair amount of tension between WMAP and SN-Gold data and one should be very careful about the priors used in these analysis. The recent SNLS data \cite{snls} is more concordant with WMAP than the SN Gold data. (b) There is no observational evidence for $w<-1$. (For more details related to these issues, see the last reference in \cite{jbp}.)
 \end{itemize}

 Given this situation, we shall now take a more serious look at the \cc\ as the source of dark energy in the universe.
 
 \section{ Cosmological Constant: Facing up to the Challenge }
 
The observational and theoretical features described above suggests that one should consider \cc\ as the most natural candidate for dark energy. Though it leads to well know fine tuning problems, it also has certain attractive features that need to kept in mind.
\begin{itemize}
\item
Cosmological constant is the most economical [just one number] and simplest  explanation for all the observations. I repeat that there is absolutely \textit{no} evidence for variation of dark energy density with redshift, which is consistent with the assumption of \cc\ .
\item
Once we invoke the \cc, classical gravity will be described by the three constants $G,c$ and $\Lambda\equiv L_\Lambda^{-2}$. It is not possible to obtain a dimensionless quantity from these; so, within classical theory, there is no fine tuning issue. Since $\Lambda(G\hbar/c^3)\equiv (L_P/L_\Lambda)^2\approx 10^{-123}$, it is obvious that the \cc\ is telling us something regarding \textit{quantum gravity}, indicated by the combination $G\hbar$. \textit{An acid test for any quantum gravity model will be its ability to explain this value;} needless to say, all the currently available models --- strings, loops etc.  --- flunk this test. Even several different approaches to semiclassical gravity \cite{semicgrav} are silent about cosmological constant.
\item
If dark energy is indeed \cc\, this will be the greatest contribution from cosmology to fundamental physics. It will be unfortunate if we miss this chance by invoking some scalar field epicycles!
\end{itemize}

In this context, it is worth  stressing another   peculiar feature of \cc\, when it is  treated as a clue to quantum gravity.
It is well known that, based on energy scales, the \cc\ problem is an infra red problem \textit{par excellence}.
At the same time, it is a relic of a quantum gravitational effect (or principle) of unknown nature. An analogy will be helpful to illustrate this point \cite{choices}. Suppose you solve the Schrodinger equation for the Helium atom for the quantum states of the two electrons $\psi(x_1,x_2)$. When the result is compared with observations, you will find that only half the states --- those in which  $\psi(x_1,x_2)$ is antisymmetric under $x_1\longleftrightarrow x_2$ interchange --- are realized in nature. But the low energy Hamiltonian for electrons in the Helium atom has no information about
this effect! Here is a low energy (IR) effect which is a relic of relativistic quantum field theory (spin-statistics theorem) that is  totally non perturbative, in the sense that writing corrections to the Helium atom Hamiltonian in some $(1/c)$ expansion will {\it not} reproduce this result. I suspect the current value of \cc\ is related to quantum gravity in a similar way. There must exist a deep principle in quantum gravity which leaves its non perturbative trace even in the low energy limit
that appears as the \cc\ (see Sections 5, 6).

Let us now turn our attention to few of the many attempts to understand the \cc\ with the choice dictated by personal bias.  A host of other approaches exist in literature, some of which can be found in \cite{catchall}.

 \subsection{ Geometrical Duality in our Universe}

 Before we discuss the ideas to explain the cosmological constant, it is important to realise some peculiar features which arise in a universe which has two independent length scales. 
 A universe with two
 length scales $L_\Lambda$ and $L_P$  will be  asymptotically De Sitter with $a(t)\propto \exp (t/L_\Lambda) $ at late times.   Given the two length scales $L_P$ and $L_\Lambda$, one can construct two energy scales
 $\rho_{_{\rm UV}}=1/L_P^4$ and $\rho_{_{\rm IR}}=1/L_\Lambda^4$ in natural units ($c=\hbar=1$). There is sufficient amount of justification from different theoretical perspectives
 to treat $L_P$ as the zero point length of spacetime \cite{zeropoint}, giving a natural interpretation to $\rho_{_{\rm UV}}$. The second one, $\rho_{_{\rm IR}}$ also has a natural interpretation. The universe which is asymptotically De Sitter has a horizon and associated thermodynamics \cite{ghds} with a  temperature
 $T=H_\Lambda/2\pi$ and the corresponding thermal energy density $\rho_{thermal}\propto T^4\propto 1/L_\Lambda^4=
 \rho_{_{\rm IR}}$. Thus $L_P$ determines the \textit{highest} possible energy density in the universe while $L_\Lambda$
 determines the {\it lowest} possible energy density in this universe. As the energy density of normal matter drops below this value, the thermal ambience of the De Sitter phase will remain constant and provide the irreducible `vacuum noise'. \textit{Note that the dark energy density is the the geometric mean $\rho_{_{\rm DE}}=\sqrt{\rho_{_{\rm IR}}\rho_{_{\rm UV}}}$ between the two energy densities.} If we define a dark energy length scale $L_{DE}$  such that $\rho_{_{\rm DE}}=1/L_{DE}^4$ then $L_{DE}=\sqrt{L_PL_\Lambda}$ is the geometric mean of the two length scales in the universe. (Incidentally, $L_{DE}\approx 0.04$ mm is macroscopic; it is also pretty close to the length scale associated with a neutrino mass of $10^{-2}$ eV; another intriguing coincidence ?!)

  \begin{figure}[ht]
  \includegraphics[angle=-90,scale=0.55]{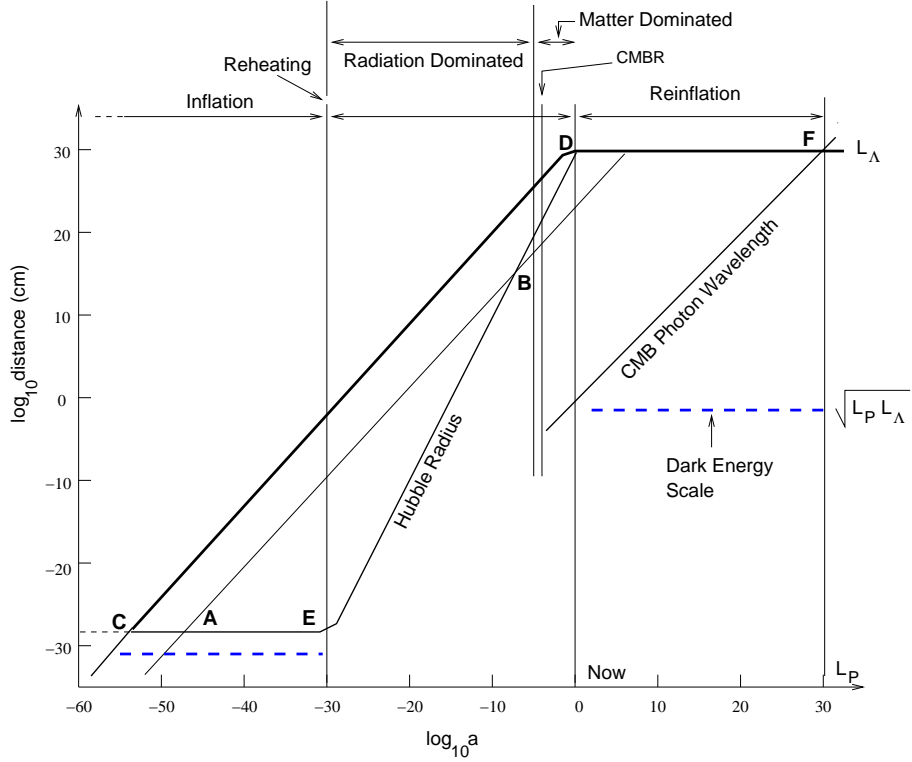}
\caption{The geometrical structure of a universe with two length scales $L_P$ and $L_\Lambda$ corresponding to the Planck length and the cosmological constant \cite{plumian,bjorken}. Such a universe spends most of its time in two De Sitter phases which are (approximately) time translation invariant. The first De Sitter phase corresponds to the inflation and the second corresponds to the accelerated expansion arising from the cosmological constant. Most of the perturbations generated during the inflation will leave the Hubble radius (at some A, say) and re-enter (at B). However, perturbations which exit the Hubble radius
earlier than C will never re-enter the Hubble radius, thereby introducing a specific  dynamic range CE during the inflationary phase. The epoch F is characterized by the redshifted CMB temperature becoming equal to the De Sitter temperature $(H_\Lambda / 2\pi)$ which introduces another dynamic range DF in the accelerated expansion after which the universe is dominated by vacuum noise
of the De Sitter spacetime.}
\label{fig:tpplumian}
  \end{figure}
 
 Figure \ref{fig:tpplumian} describes some peculiar features in such a universe \cite{plumian,bjorken}. Using the characteristic length scale of expansion,
 the Hubble radius $d_H\equiv (\dot a/a)^{-1}$, we can distinguish between three different phases of such a universe. The first phase is when the universe went through a inflationary expansion with $d_H=$ constant; the second phase is the radiation/matter dominated phase in which most of the standard cosmology operates and $d_H$ increases monotonically; the third phase is that of re-inflation (or accelerated expansion) governed by the cosmological constant in which $d_H$ is again a constant. The first and last phases are time translation invariant;
 that is, $t\to t+$ constant is an (approximate) invariance for the universe in these two phases. The universe satisfies the perfect cosmological principle and is in steady state during these phases!
 
 In fact, one can easily imagine a scenario in which the two deSitter phases (first and last) are of arbitrarily long duration \cite{plumian}. If  $\Omega_\Lambda\approx 0.7, \Omega_{DM}\approx 0.3$ the final deSitter phase \textit{does} last forever; as regards the inflationary phase, nothing prevents it from lasting for arbitrarily long duration. Viewed from this perspective, the in between phase --- in which most of the `interesting' cosmological phenomena occur ---  is  of negligible measure in the span of time. It merely connects two steady state phases of the universe.
 The figure \ref{fig:tpplumian} also shows the variation of $L_{DE}$ by broken horizontal lines. 
 
 While the two deSitter phases can last forever in principle, there is a natural cut off length scale in both of them
 which makes the region of physical relevance to be finite \cite{plumian}. Let us first discuss the case of re-inflation in the late universe. 
 As the universe grows exponentially in the phase 3, the wavelength of CMBR photons are being redshifted rapidly. When the temperature of the CMBR radiation drops below the deSitter temperature (which happens when the wavelength of the typical CMBR photon is stretched to the $L_\Lambda$.)
 the universe will be essentially dominated by the vacuum thermal noise of the deSitter phase.
 This happens at the point marked F when the expansion factor is $a=a_F$ determined by the
  equation $T_0 (a_0/a_{F}) = (1/2\pi L_\Lambda)$. Let $a=a_\Lambda$ be the epoch at which
  cosmological constant started dominating over matter, so that $(a_\Lambda/a_0)^3=
  (\Omega_{DM}/\Omega_\Lambda)$. Then we find that the dynamic range of 
 DF is 
 \begin{equation}
\frac{a_F}{a_\Lambda} = 2\pi T_0 L_\Lambda \left( \frac{\Omega_\Lambda}{\Omega_{DM}}\right)^{1/3}
\approx 3\times 10^{30}
\end{equation}

 Interestingly enough, one can also impose a similar bound on the physically relevant duration of inflation. 
 We know that the quantum fluctuations generated during this inflationary phase could act as seeds of structure formation in the universe \cite{genofpert}. Consider a perturbation at some given wavelength scale which is stretched with the expansion of the universe as $\lambda\propto a(t)$.
 (See the line marked AB in Figure \ref{fig:tpplumian}.)
 During the inflationary phase, the Hubble radius remains constant while the wavelength increases, so that the perturbation will `exit' the Hubble radius at some time (the point A in Figure \ref{fig:tpplumian}). In the radiation dominated phase, the Hubble radius $d_H\propto t\propto a^2$ grows faster than the wavelength $ \lambda\propto a(t)$. Hence, normally, the perturbation will `re-enter' the Hubble radius at some time (the point B in Figure \ref{fig:tpplumian}).
 If there was no re-inflation, this will make {\it all} wavelengths re-enter the Hubble radius sooner or later.
 But if the universe undergoes re-inflation, then the Hubble radius `flattens out' at late times and some of the perturbations will {\it never} reenter the Hubble radius! The limiting perturbation which just `grazes' the Hubble radius as the universe enters the re-inflationary phase is shown by the line marked CD in Figure \ref{fig:tpplumian}. If we use the criterion that we need the perturbation to reenter the Hubble radius, we get a natural bound on the duration of inflation which is of direct astrophysical relevance. This portion of the inflationary regime is marked by CE
 and can be calculated as follows: Consider  a perturbation which leaves the Hubble radius ($H_{in}^{-1}$) during the inflationary epoch at $a= a_i$. It will grow to the size $H_{in}^{-1}(a/a_i)$ at a later epoch. 
 We want to determine $a_i$ such that this length scale grows to 
   $L_\Lambda$ just when the dark energy starts dominating over matter; that is at
 the epoch $a=a_\Lambda = a_0(\Omega_{DM}/\Omega_{\Lambda})^{1/3}$. 
  This gives 
  $H_{in}^{-1}(a_\Lambda/a_i)=L_\Lambda$ so that $a_i=(H_{in}^{-1}/L_\Lambda)(\Omega_{DM}/\Omega_{\Lambda})^{1/3}a_0$. On the other hand, the inflation ends at 
  $a=a_{end}$ where $a_{end}/a_0 = T_0/T_{\rm reheat}$ where $T_{\rm reheat} $ is the temperature to which the universe has been reheated at the end of inflation. Using these two results we can determine the dynamic range of CE to be 
  \begin{equation}
\frac{a_{\rm end} }{a_i} = \left( \frac{T_0 L_\Lambda}{T_{\rm reheat} H_{in}^{-1}}\right)
\left( \frac{\Omega_\Lambda}{\Omega_{DM}}\right)^{1/3}=\frac{(a_F/a_\Lambda)}{2\pi T_{\rm reheat} H_{in}^{-1}} \cong 10^{25}
\end{equation} 
where we have used the fact that, for a GUTs scale inflation with $E_{GUT}=10^{14} GeV,T_{\rm reheat}=E_{GUT},\rho_{in}=E_{GUT}^4$
we have $2\pi H^{-1}_{in}T_{\rm reheat}=(3\pi/2)^{1/2}(E_P/E_{GUT})\approx 10^5$.
If we consider a quantum gravitational, Planck scale, inflation with $2\pi H_{in}^{-1} T_{\rm reheat} = \mathcal{O} (1)$, the phases CE and DF are approximately equal. The region in the quadrilateral CEDF is the most relevant part of standard cosmology, though the evolution of the universe can extend to arbitrarily large stretches in both directions in time. 

This figure is definitely telling us something regarding the duality between Planck scale and Hubble scale or between the infrared and ultraviolet limits of the theory.
The mystery is compounded by the fact the asymptotic de Sitter phase has an observer dependent horizon and
related thermal properties. Recently, it has been shown --- in a series of papers, see ref.\cite{tpholo} ---  that it is possible to obtain 
classical relativity from purely thermodynamic considerations.  It is difficult to imagine that these features are unconnected and accidental; at the same time, it is difficult to prove a definite connection between these ideas and the \cc. I will say more about this in Sections 5, 6.

\section{ Attempts on the cosmological constant's life}

\subsection{ Dark energy from a nonlinear correction term}

One of the \textit{least} esoteric ideas regarding the dark energy
is that the cosmological constant term in the FRW equations arises because we have not calculated the energy density driving the expansion of the universe correctly. The motivation for such a suggestion arises from the following fact:  The energy momentum tensor of the real universe, $T_{ab}(t,{\bf x})$ is inhomogeneous and anisotropic and will lead to a very complex metric $g_{ab}$ if only we could solve the exact Einstein's equations
$G_{ab}[g]=\kappa T_{ab}$.
The metric describing the large scale structure of the universe should be obtained by averaging this exact solution over a large enough scale, to get $\langle g_{ab}\rangle $. But what we actually do is to average the stress tensor {\it first} to get $\langle T_{ab}\rangle $ and {\it then} solve Einstein's equations. But since $G_{ab}[g]$ is  nonlinear function of the metric, $\langle G_{ab}[g]\rangle \neq G_{ab}[\langle g\rangle ]$ and there is a discrepancy. This is most easily seen by writing
\begin{equation}
G_{ab}[\langle g\rangle ]=\kappa [\langle T_{ab}\rangle  + \kappa^{-1}(G_{ab}[\langle g\rangle ]-\langle G_{ab}[g]\rangle )]\equiv \kappa [\langle T_{ab}\rangle  + T_{ab}^{corr}]
\end{equation}
If --- based on observations --- we take the $\langle g_{ab}\rangle $ to be the standard Friedman metric, this equation shows that it has, as its  source,  \textit{two} terms:
The first is the standard average stress tensor and the second is a purely geometrical correction term
$T_{ab}^{corr}=\kappa^{-1}(G_{ab}[\langle g\rangle ]-\langle G_{ab}[g]\rangle )$ which arises because of nonlinearities in the Einstein's theory that  leads to $\langle G_{ab}[g]\rangle \neq G_{ab}[\langle g\rangle ]$. If this term can mimic the \cc\ at large scales there will be no need for dark energy and --- as a bonus --- one will solve the coincidence problem!
 This effect, of course, is real (for an explicit example, in a different context of electromagnetic plane wave, see \cite{gofemw}) but is usually quite small.  In spite of some recent attention this idea has received \cite{flucde} I doubt whether the idea will lead to the correct result when implemented properly.  In a way, the problem is to average physically positive quantities and obtain a result which is not only negative but is sufficiently negative to dominate over
 positive matter density. One possible way to attack the nonlinear back reaction is to use some analytic approximations to nonlinear perturbations (usually called non-linear scaling relations, see e.g. \cite{nsr}) to estimate this term.
This does not lead to a stress tensor that mimics dark energy (Padmanabhan, unpublished). 

\subsection{ Cosmic Lenz law}

The second simplest possibility which has been attempted in the literature several times in different guises is to try and  ''cancel out'' the \cc\ by some process,
usually quantum mechanical in origin. One can, for example, ask whether switching on a \cc\ will
lead to a vacuum polarization with an effective energy momentum tensor that will tend to cancel out the \cc.
A less subtle way of doing this is to invoke another scalar field (here we go again!) such that it can couple to 
\cc\ and reduce its effective value \cite{lenz}. Unfortunately, none of this could be made to work properly. By and large, these approaches lead to an energy density which is either $\rho_{_{\rm UV}}\propto L_P^{-4}$ (where 
 $L_P$ is the Planck length) or to $\rho_{_{\rm IR}}\propto L_\Lambda^{-4}$ (where 
 $L_\Lambda=H_\Lambda^{-1}$ is the Hubble radius associated with the \cc\ ). The first one is too large while the second one is too small! 

\subsection{ Unimodular Gravity}

One possible way of addressing the issue of \cc\ is to simply eliminate from the gravitational theory those modes which couple to cosmological constant. If, for example, we have a theory in which the source of gravity is
$(\rho +p)$ rather than $(\rho +3p)$ in Eq. (\ref{nextnine}), then \cc\ will not couple to gravity at all. (The non linear coupling of matter with gravity has several subtleties; see eg. \cite{gravitonmyth}.) Unfortunately
it is not possible to develop a covariant theory of gravity using $(\rho +p)$ as the source. But we can probably gain some insight from the following considerations. Any metric $g_{ab}$ can be expressed in the form $g_{ab}=f^2(x)q_{ab}$ such that
${\rm det}\, q=1$ so that ${\rm det}\, g=f^4$. From the action functional for gravity
\begin{equation}
A=\frac{1}{16\pi G}\int \sqrt{-g}\, d^4x\,  (R -2\Lambda)
=\frac{1}{16\pi G}\int \sqrt{-g} \, d^4x\,  R -\frac{\Lambda}{8\pi G}\int d^4x f^4(x)
\label{oneone}
\end{equation}
it is obvious that the \cc\ couples {\it only} to the conformal factor $f$. So if we consider a theory of gravity in which $f^4=\sqrt{-g}$ is kept constant and only $q_{ab}$ is varied, then such a model will be oblivious of
direct coupling to \cc. If the action (without the $\Lambda$ term) is varied, keeping ${\rm det}\, g=-1$, say, then one is lead to a {\it unimodular theory of gravity} that has  the equations of motion 
$R_{ab}-(1/4)g_{ab}R=\kappa(T_{ab}-(1/4)g_{ab}T)$ with zero trace on both sides. Using the Bianchi identity, it is now easy to show that this is equivalent to the usual  theory with an {\it  arbitrary} \cc. That is, \cc\ arises as an undetermined integration constant in this model \cite{unimod}. 
(The same result arises in another, completely different approach to gravity which we will discuss in the next section.)

While this is all very interesting, we still need an extra physical principle to fix the value (even the sign) of \cc\ .
One possible way of doing this, suggested by \eq{oneone}, is to  interpret the $\Lambda$ term in the action as a Lagrange multiplier for the proper volume of the spacetime. Then it is reasonable to choose the \cc\ such that the total proper volume of the universe is equal to a specified number. While this will lead to a \cc\ which has the correct order of magnitude, it has an  obvious problem because  the proper four volume of the universe is infinite unless we make the spatial sections compact and restrict the range of time integration. 

\section{Gravity from the surface degrees of freedom and the cosmological constant}

The  failure of simple ideas suggests that the problem of \cc\ will not allow any `quick fix' solution and is possibly deeply entrenched in quantum gravity. We have got into trouble because of our naive expectations related to how gravity couples to the vacuum modes.
Hence it is necessary to take a few steps back and review all our ideas related to this.

The issue of \cc\ is intimately related to two nontrivial questions: (i) How is the \textit{microscopic} structure of the vacuum state modified by gravity ? (ii) What kind of \textit{macroscopic} gravitational field is produced by the vacuum ? It is, for example, well known that the vacuum state  depends on the class of observers we are considering \cite{probe}. In \textit{any} spacetime, there will exist families of observers (congruence of timelike curves) 
who will have access to only part of the spacetime. 
The well known examples are observers at $r=$ constant$>2M$
in the Schwarzschild spacetime or the uniformly accelerated observers in flat spacetime.
Any class of observers, of course, has an equal right to describe physical phenomena entirely in terms of the variables defined in the regions accessible to them. 
The action
functional describing gravity, used by these observers (who have access to only part of the spacetime) should depend only on the variables defined on the region accessible to them, including the boundary of this region. (This is essentially the philosophy of renormalization group theory translated 
from momentum space  into real space.) Since the horizon and associated boundaries may exist for some observers  but not for others, this brings up a new level of observer dependence in the action functional describing the theory, though the existence of horizons.  \textit{The physics of the region blocked by the horizon will be then encoded in a boundary term in the action. }

In fact, it is possible to obtain (see the first paper in Ref. \cite{tpholo}) the dynamics of gravity from an approach which  uses \textit{only} the surface term of the Hilbert action; \textit{ we do not need the bulk term at all}!.
In this approach, the action functional for the continuum spacetime is
\begin{equation}
A_{tot}=A_{sur}+A_{matter}=
\frac{1}{16\pi G}\int_{\partial\mathcal{V}} d^3 x \, 
\sqrt{-g}n_cQ_a^{\phantom{a}bcd}\Gamma^a_{bd}
+\int_{\mathcal{V}} d^4x \, \sqrt{-g}L_{m}(g,\phi)
\label{actfunc}
\end{equation}
where $Q_a{}^{bcd}=(1/2)(-\delta^c_ag^{bd}+\delta^d_ag^{bc})$.
Matter degrees of freedom live in the bulk $\mathcal{V}$ while the gravity contributes on the boundary ${\partial\mathcal{V}}$. When the boundary
has a part which acts as a horizon for a class of observers, 
we demand that the action should be invariant under virtual displacements of this horizon. \textit{This leads to Einstein's theory with a cosmological constant that arises as an integration constant} \cite{tpholo}. We will call this the holographic dual description of standard Einstein gravity. 
(The term holography is used by different people in different contexts; when I use that term in this talk, I use it in the sense described above.)

What is more important is that, in this approach, gravity has a thermodynamic interpretation.
The $A_{sur}$ is directly related to the (observer dependent horizon) entropy and
 its variation, when the horizon is moved infinitesimally, is equivalent to the change in the entropy $dS$ due to virtual work. The variation
 of the matter term contributes the $PdV$ and $dE$ terms and the entire variational principle is equivalent to the thermodynamic identity 
 \begin{equation}
 TdS=dE+PdV
 \end{equation}
 applied to the changes when a horizon undergoes a virtual displacement. In the case of spherically symmetric spacetimes, for example, it can be
\textit{explicitly} demonstrated \cite{ss} that the Einstein's equations follow from the thermodynamic
identity applied to horizon displacements. 

In this approach, the 
 continuum spacetime is like an elastic solid (`Sakharov paradigm'; see e.g. \cite{sakharov}) with Einstein's equations providing the macroscopic description. 
 In the virtual displacement $x^a\to \bar{x}^a=x^a+\xi^a$ the
$\xi^a(x)$ is similar to the displacement vector field used, for example,
in the study of elastic solids. 
The true degrees of freedom are some unknown `atoms of spacetime' but in the continuum limit,
the displacement $x^a\to \bar{x}^a=x^a+\xi^a(x)$ captures the relevant dynamics,  just like 
in the study of elastic properties of the continuum solid. Further, it can be shown that the horizons in the spacetime are  similar to defects in the solid so that their displacement costs entropy. 
This suggests that \textit{the true  degrees of freedom of gravity
for a volume $\mathcal{V}$
reside in its boundary $\partial\mathcal{V}$} --- a  point of view that is strongly supported by the study
of horizon entropy, which shows that the degrees of freedom hidden by a horizon scales as the area and not as the volume. The \cc\ arises  as a undetermined integration constant but closely related to the `bulk expansion' of the solid.

There is actually a deep reason as to why this works, which actually goes beyond the Einstein's theory. Similar results exists for \textit{any theory based on principle of equivalence}, in which the gravity is described by
a metric tensor $g_{ab}$. Let me briefly describe the general setting from which this thermodynamic picture arises \cite{qn}.

Consider a  (generalized) theory of gravity in D-dimensions based on a generally covariant
scalar lagrangian $L$ which is a functional of the metric $g^{ab}$ and curvature $R^a_{\phantom{a}bcd}$.
Instead of treating $[g^{ab},\partial_cg^{ab},
\partial_d\partial_cg^{ab}$] as the independent variables,  it is convenient to use $[g^{ab},
\Gamma^i_{kl},R^a_{\phantom{a}bcd}]$ as the independent variables. The curvature tensor $R^a_{\phantom{a}bcd}$ can be expressed entirely in terms of $\Gamma^i_{kl}$ and $\partial_j\Gamma^i_{kl}$ and is \textit{independent} of $g^{ab}$. It is also useful to define the tensor $P_a^{\phantom{a}bcd}\equiv (\partial L/\partial R^a_{\phantom{a}bcd} )$
which has exactly the same symmetries of $R^a_{\phantom{a}bcd}$. Varying  the action functional 
gives
\begin{equation}
\delta A = \delta \int_\mathcal{V} d^Dx\sqrt{-g}\, L = \int_\mathcal{V} d^Dx \, \sqrt{-g} \, E_{ab} \delta g^{ab} + 
\int_\mathcal{V} d^Dx \, \sqrt{-g} \, \nabla_j \delta v^j
\label{deltaa}
\end{equation}
where
\begin{equation}
E_{ab}\equiv\left( \frac{\partial \sqrt{-g}L}{\partial g^{ab}} 
-2\sqrt{-g}\nabla^m\nabla^n P_{amnb} \right)
\label{eab}
\end{equation} 
and
\begin{equation}
\delta v^j \equiv [2P^{ibjd}(\nabla_b\delta g_{di})-2\delta g_{di}(\nabla_cP^{ijcd})]
\label{genvc}
\end{equation} 
This result is completely general. In $\delta A$ in \eq{deltaa}, the second term will lead to a surface contribution. 
To have a good variational principle leading to the result
$E^{ab} = $ matter source terms, we need to assume that $n_a \delta v^a =0$ on 
$\partial \mathcal{V}$ where $n_a$ is the normal to the boundary. In general this requires a particular combination of the ``coordinates" [$g_{ab}$] and the ``momenta" [$\nabla_c\delta g_{ab}$] to vanish and we need to put conditions on both the dynamical variables \textit{and} their derivatives on the boundary.  
It is more reasonable in a quantum theory to choose either the variations of coordinates or those of momenta to vanish rather than a linear combination.  To achieve this, let us concentrate on a subset of Lagrangians for which $P^{abcd}$
is divergence free. That is, we demand:
\begin{equation}
\nabla_cP^{ijcd}=0
\label{condition}
\end{equation} 
Because of the symmetries, this means $P^{abcd}$ is divergence-free in \textit{all} indices. Then
\eq{eab} shows that  the source-free equations of motion $E_{ab}=0$ reduces to 
\begin{equation}
\frac{\partial \sqrt{-g}L}{\partial g^{ab}}=0
\label{ordder}
\end{equation} 
That is, just setting the \textit{ordinary derivative} of lagrangian density with respect to $g^{ab}$ to zero will give the equations of motion!. 

Interestingly enough, this condition encompasses \textit{all} the gravitational theories (in D dimensions) in which
the field equations are no higher than second degree, though we did \textit{not }demand that explicitly. To see this, let us consider the possible fourth rank tensors $P^{abcd}$ which (i) have the symmetries of curvature tensor; (b) are divergence-free; (iii) made from $g^{ab}$ and  $R^a_{\phantom{a}bcd}$. If we do not use the curvature tensor, then we have just one choice
 made from metric:
\begin{equation}
P_a^{\phantom{a}bcd}=\frac{1}{2}(\delta^c_ag^{bd}-\delta^d_ag^{bc})
\label{pforeh}
\end{equation} 
This leads to the Einstein-Hilbert action
\begin{equation}
L\equiv P_a^{\phantom{a}bcd}R^a_{\phantom{a}bcd}+{\rm constant}\equiv R-2\Lambda
\end{equation} 
with an integration constant which is the cosmological constant. It is easy to verify that the standard Einstein's field equations arise from the \textit{ordinary derivative}
taken in \eq{ordder} with the variables as we have chosen: $\sqrt{-g}L=\sqrt{-g}[g^{ab}R_{ab}-2\Lambda]$. 

Next, if we allow for $P_a^{\phantom{a}bcd}$ to depend linearly on curvature, then we have the following 
additional choice of  tensor with required symmetries:
\begin{equation}
P^{abcd}=R^{abcd} -  G^{ac}g^{bd}+ G^{bc}g^{ad} + R^{ad}g^{bc} - R^{bd}g^{ac} 
\label{ping}
\end{equation} 
(In four dimensions, this tensor is essentially the double-dual of
$R_{abcd}$ and in any dimension can be obtained from $R_{abcd}$ using the alternating tensor \cite{gliner}.)
In this case, integrating
$(\partial L/\partial R^a_{\phantom{a}bcd})=P_a^{\phantom{a}bcd}$, we get
\begin{equation}
L=\frac{1}{2}\left(g_{ia}g^{bj}g^{ck}g^{dl}-4g_{ia}g^{bd}g^{ck}g^{jl}
+\delta^c_a\delta^k_ig^{bd}g^{jl}\right)R^i_{\phantom{i}jkl}R^a_{\phantom{a}bcd}
=\frac{1}{2}\left[R^{abcd}R_{abcd}-4R^{ab}R_{ab}+R^2\right]
\end{equation} 
(plus, of course, the cosmological constant which we have not exhibited). This just the Gauss-Bonnet(GB) action which is a pure divergence in 4 dimensions but not in higher dimensions.
The unified procedure for deriving Einstein-Hilbert action and GB action [essentially from the condition in \eq{condition}] shows that they are more closely related  to each other than previously suspected. The fact that \textit{several string theoretical models get GB type terms as corrections} is noteworthy in this regard. 

In fact, the holographic dual description of gravity --- in which the same equations arise from a surface term --- exists for all these theories. To obtain this,  note that any nontrivial scalar lagrangian built from $R^a_{\phantom{a}bcd}$ can be written in the form
$
L=Q_a^{\phantom{a}bcd}R^a_{\phantom{a}bcd}
$ 
without  loss of generality. (The tensor $Q_a^{\phantom{a}bcd}$ depends on curvature and metric; so we have not made any restrictive assumptions \cite{clarify}).  Using the antisymmetry of $Q_a^{\phantom{a}bcd}$ in $c,d$ we can write:
\begin{equation}
\sqrt{-g}L=\sqrt{-g}Q_a^{\phantom{a}bcd}R^a_{\phantom{a}bcd}
=2\sqrt{-g}Q_a^{\phantom{a}bcd}[\partial_c\Gamma^a_{db}+\Gamma^a_{ck}\Gamma^k_{db}]
\end{equation}  We now do an integration by parts, use
$\partial_c\ln \sqrt{-g}=\Gamma^k_{ck}$ and express $\partial_cQ_a^{\phantom{a}bcd}$ in terms of $\nabla_cQ_a^{\phantom{a}bcd}$
and four terms of type $\Gamma Q$. This gives a remarkably simple result:
\begin{equation}
\sqrt{-g}L=2\partial_c\left[\sqrt{-g}Q_a^{\phantom{a}bcd}\Gamma^a_{bd}\right]
+2\sqrt{-g}Q_a^{\phantom{a}bcd}\Gamma^a_{dj}\Gamma^j_{bc}
 -2\sqrt{-g}\Gamma^a_{bd}\nabla_cQ_a^{\phantom{a}bcd}
 \label{simple}
\end{equation} 
So far we have not made any assumptions and this result shows that any scalar gravitational lagrangian
built from metric and curvature has a separation into a surface term (first term) and bulk term (second and third terms) in a natural but non covariant manner.  Ignoring the surface term, one can obtain the same \textit{covariant} equations of motion as before even from a \textit{non covariant} lagrangian.
In general the equations of motion will be higher than second order but let us again specialize to the case in which  $Q_a^{\phantom{a}bcd}$ is divergence-free. Then the last term \eq{simple} vanishes and we get the simple result:
\begin{equation}
\sqrt{-g}L=2\partial_c\left[\sqrt{-g}Q_a^{\phantom{a}bcd}\Gamma^a_{bd}\right]
+2\sqrt{-g}Q_a^{\phantom{a}bcd}\Gamma^a_{dk}\Gamma^k_{bc}\equiv L_{\rm sur} + L_{\rm bulk} 
\label{gensq}
 \end{equation} 
which --- in particular --- should hold for \textit{both} Einstein-Hilbert and GB actions. 
The second term is the generalisation of the standard $\Gamma^2$ action for GB action. 
In the case of both Einstein-Hilbert action and 
GB action one can take $Q_a^{\phantom{a}bcd} = P_a^{\phantom{a}bcd}$ with suitable normalization.
 When $Q_a^{\phantom{a}bcd}$ is built from metric alone, it is given by \eq{pforeh} and \eq{gensq} becomes
\begin{equation}
\sqrt{-g}L=\partial_c\left[\sqrt{-g}(g^{bd}\Gamma_{bd}^c-g^{bc}\Gamma^a_{ba})\right]
+\sqrt{-g}(g^{bd}\Gamma^a_{dj}\Gamma^j_{ba}-g^{bc}\Gamma^a_{aj}\Gamma^j_{bc})
\label{surbulkeh}
\end{equation} 
which is precisely the bulk-surface decomposition for Einstein-Hilbert action. 
In the case of GB action, we get a similar result with $Q_a^{\phantom{a}bcd}$ being
given by the right hand side of \eq{ping}. 

There is also another striking relation between the surface and bulk terms in the lagrangian
in \eq{gensq}. To see this we 
  begin by noting
that, the two parts:
\begin{equation}
L_{\rm bulk} = 2 \sqrt{-g}\, Q_a^{\phantom{a}bcd} \Gamma^a_{dk} \Gamma^k_{bc}; \qquad
L_{\rm sur}=2\partial_c\left[\sqrt{-g}Q_a^{\phantom{a}bcd}\Gamma^a_{bd}\right]
\equiv\partial_c[\sqrt{-g}V^c]
\end{equation} 
 are both contain the same information in terms of $Q_a^{\phantom{a}bcd}$ and hence could be \textit{always} related to each other. It is easy to verify \cite{note1} that 
 \begin{equation}
L_{\rm sur} = -\frac{1}{2} \partial_c \left(\delta^k_b \frac{\partial L_{\rm bulk}}{\partial \Gamma^k_{cb}}
\right)_{g,R}
\label{holo}
\end{equation} 
In the case of Einstein-Hilbert action, there is a still simpler relation:
 \begin{equation}
[(D/2)-1]L_{\rm sur} =  -\partial_a \left(g_{bc}\frac{\partial L_{\rm bulk}}{\partial(\partial_a g_{bc})}\right)
\end{equation} 
(For the physical significance of this structure, see the papers in \cite{tpholo}.)

Finally, probably truer to the term of holography, there is a relation that determines the form of $L_{bulk}$ and $L$
if we know the form of $L_{sur}$ or --- equivalently --- the $V^c$. We have:
\begin{equation}
L=\frac{1}{2}R^a_{\phantom{a}bcd}\left(\frac{\partial V^c}{\partial \Gamma^a_{bd}}\right)_{g,R};
\qquad
L_{bulk}=\sqrt{-g}\left(\frac{\partial V^c}{\partial \Gamma^a_{bd}}\right)_{g,R}
\Gamma^a_{dk}\Gamma^k_{bc}
\end{equation} 
The first relation also shows that  $(\partial V^c/\partial \Gamma^a_{bd})$ is generally covariant in spite of the appearance. This realtion makes
the action intrinsically holographic with the surface term containing an equivalent 
information as the bulk. 

There are two comments worth making about the above derivation: First, much of the index gymnastics can be eliminated by introducing
a set of tetrads $e^{\,c}_{(k)}$ where $k=0,1, ..., D$ identifies the vector and $c$
indicates the component. The dual basis is given by  $e_d^{(k)}$ with
$e_d^{(k)} e^{\,c}_{(k)} = \delta^c_d$.  Writing 
$R^{\,c}_{\phantom{\,c}dba} = e_d^{(k)} \nabla_{[b} \nabla_{a]}\,  e_{(k)}^{\,c}$ 
our Lagrangian becomes 
\begin{equation}
L = Q_c^{\phantom{c}dba}\,  R^{\,c}_{\phantom{\,c}dba} = 2 Q_c^{\phantom{c}dba}\,  e_d^{(k)}\, \nabla_{b} \nabla_{a}\,  e^{\,c}_{(k)}
= \nabla_b \left( 2 Q_c^{\phantom{c}dba}\,  e_d^{(k)}\, \nabla_a \, e^{\,c}_{(k)}\right)- 2 Q_c^{\phantom{c}dba}
\left( \nabla_b\,  e^{(k)}_d\right) \left( \nabla_a \, e^{\,c}_{(k)}\right)
\end{equation}
where we have done an integration by parts and used $\nabla_b Q_c^{\phantom{c}dba} =0$.
This reduces to \eq{gensq} in a coordinate basis with 
$e^c_{(k)} = \delta^c_k,\nabla_a \, e^{\,c}_{(k)}=\Gamma^c_{ak},\nabla_b\,  e^{(k)}_d=-\Gamma^k_{bd}$.
One can also express \eq{holo} in a similar manner. 

Second, note that \eq{gensq} with $\nabla_b Q_a^{\phantom{a}bcd}=0$ represents the most general effective lagrangian for gravity which is consistent with principle of equivalence, general covariance and the dynamical requirement that a well-defined variational principle should exist. The structure of the theory is specified by a single divergence-free fourth rank tensor $Q_a^{\phantom{a}bcd}$ having the symmetries of the curvature tensor. The semi classical,
low energy, action for gravity can now be determined from the derivative expansion
of $Q_a^{\phantom{a}bcd}$ in powers of number of derivatives: 
\begin{equation}
Q_a^{\phantom{a}bcd} (g,R) = \overset{(0)}{Q}_a{}^{bcd} (g) + \alpha\, \overset{(1)}{Q}_a{}^{bcd} (g,R) + \beta\, \overset{(2)}{Q}_a{}^{bcd} (g,R^2,\nabla R) + \cdots
\label{derexp}
\end{equation} 
where $\alpha, \beta, \cdots$ are coupling constants. The first term gives Einstein-Hilbert action and the second one is Gauss-Bonnet action. It is worth recalling \textit{that such a Gauss-Bonnet term
arises as the correction in string theories} \cite{zw}, as to be expected from our general principle. 
It is also remarkable that any such Lagrangian 
$L=Q_a^{\phantom{a}bcd}R^a_{\phantom{a}bcd}$
with $\nabla_b\,Q_a^{\phantom{a}bcd}=0$ can be decomposed into a surface and bulk terms which are
related holographically. 
(The lagrangian in gauge theories also has similar structure and one can repeat most of the above analysis \cite{puri}. This leads to an interesting
relationship between gauge theories and gravity  though --- as is well known --- the lack of a metric leads to significant structural dfifferences.)

Everything else goes through as before [in this case, when $\nabla_a Q^{abcd}=0$] and it is possible
to reformulate the theory retaining \textit{only} the surface term for the gravity sector as in the
case of Einstein gravity. [For a related but alternative approach, see \cite{thomas}].
 If one considers the infinitesimal virtual displacement $x^a\to x^a+\xi^a$ of the horizon, and use the fact that any scalar density changes by $\delta (\sqrt{-g}S)=-\sqrt{-g}\nabla_a(S\xi^a)$ one can 
 two key results: First is the identity 
$
\nabla_a E^{ab}=0
$
which is just the generalization of Bianchi identity. Second, if we consider an action principle with
based on $(A_m+A_s)$ where $A_m$ is the matter action and $A_s$ is the action obtained from $-L_{sur}$
(the minus sign is just to ensure that this is the term which, when \textit{added} to our action will \textit{cancel} the surface term) then, for variations that arise from displacement of a horizon normal to itself, one gets the equation $(E_{ab}-\frac{1}{2}T_{ab})\xi^b\xi^a=0$ where $\xi^a$ is \textit{null}. Combined with  identity $\nabla_a E^{ab}=0$ this will lead to standard field equations with a cosmological term $E_{ab}=(1/2)T_{ab}+\Lambda g_{ab}$ just as in the case of Einstein-Hilbert action (derived by this route in the first two papers in \cite{tpholo}). 

Once again \textit{the cosmological constant arises as an integration constant}.
There are three key morals to this story: 
\begin{itemize}
\item
First, the bulk degrees of freedom are gauge redundant and what is really important are the surface degrees of freedom. The bulk \cc\ which is purely an integration constant should not play any observable role. 
\item
Second,  If an observer has a horizon, we should take that seriously, and work with the degrees of freedom confined by the horizon. This clearly changes the pattern of vacuum \textit{fluctuations}.
\item
Third, and most important, these features arise purely from principle of equivalence and general covariance and is not specific to Einstein's theory. Any theory that has a metric description will have similar features and hence \textit{higher order quantum gravitational corrections are likely to obey these principles.}
\end{itemize}
I will now show how these ideas lead to a workable model for \cc.

\section{ A model that works: Gravity as detector of the vacuum fluctuations}

Finally, I will describe an idea which \textit{does} lead to the correct value of \cc\ which is based on the following three key ingredients:
\begin{itemize}
\item
The description of gravity based on purely a surface term in the action provides a natural back drop for ignoring the bulk value of the cosmological constant. This is consistent with the fact that in this approach, bulk \cc\ arises as an integration constant. 
\item
What is observable through gravitational effects, in the correct theory of quantum gravity, should be the \textit{fluctuations} in the vacuum energy and \textit{not} the absolute value of the vacuum energy.
\item
These fluctuations will be non-zero if the universe has a deSitter horizon which provides a confining 
volume.
\end{itemize}
Let me now elaborate on this idea. The conventional discussion of the relation between cosmological constant and vacuum energy density is based on
evaluating the zero point energy of quantum fields with an ultraviolet cutoff and using the result as a 
source of gravity.
Any reasonable cutoff will lead to a vacuum energy density $\rho_{\rm vac}$ which is unacceptably high. 
This argument,
however, is too simplistic since the zero point energy --- obtained by summing over the
$(1/2)\hbar \omega_k$ --- has no observable consequence in any other phenomena and can be subtracted out by redefining the Hamiltonian. The observed non trivial features of the vacuum state of QED, for example, arise from the {\it fluctuations} (or modifications) of this vacuum energy rather than the vacuum energy itself. 
This was, in fact,  known fairly early in the history of cosmological constant problem and, in fact, is stressed by Zeldovich \cite{zeldo} who explicitly calculated one possible contribution to {\it fluctuations} after subtracting away the mean value.
This
suggests that we should consider   the fluctuations in the vacuum energy density in addressing the 
cosmological constant problem. 

If the vacuum probed by the gravity can readjust to take away the bulk energy density $\rho_{_{\rm UV}}\simeq L_P^{-4}$, quantum \textit{fluctuations} can generate
the observed value $\rho_{\rm DE}$. One of the simplest models \cite{tpcqglamda} which achieves this uses the fact that, in the semiclassical limit, the wave function describing the universe of proper four-volume ${\cal V}$ will vary as
$\Psi\propto \exp(-iA_0) \propto 
 \exp[ -i(\Lambda_{\rm eff}\mathcal V/ L_P^2)]$. If we treat 
  $(\Lambda/L_P^2,{\cal V})$ as conjugate variables then uncertainty principle suggests $\Delta\Lambda\approx L_P^2/\Delta{\cal V}$. If
the four volume is built out of Planck scale substructures, giving $ {\cal V}=NL_P^4$, then the Poisson fluctuations will lead to $\Delta{\cal V}\approx \sqrt{\cal V} L_P^2$ giving
    $ \Delta\Lambda=L_P^2/ \Delta{\mathcal V}\approx1/\sqrt{{\mathcal V}}\approx   H_0^2
 $. (This idea can be a more quantitative; see \cite{tpcqglamda}).

Similar viewpoint arises, more formally, when we study the question of \emph{detecting} the energy
density using gravitational field as a probe.
 Recall that an Unruh-DeWitt detector with a local coupling $L_I=M(\tau)\phi[x(\tau)]$ to the {\it field} $\phi$
actually responds to $\langle 0|\phi(x)\phi(y)|0\rangle$ rather than to the field itself \cite{probe}. Similarly, one can use the gravitational field as a natural ``detector" of energy momentum tensor $T_{ab}$ with the standard coupling $L=\kappa h_{ab}T^{ab}$. Such a model was analysed in detail in ref.~\cite{tptptmunu} and it was shown that the gravitational field responds to the two point function $\langle 0|T_{ab}(x)T_{cd}(y)|0\rangle $. In fact, it is essentially this fluctuations in the energy density which is computed in the inflationary models \cite{inflation} as the seed {\it source} for gravitational field, as stressed in
ref.~\cite{tplp}. All these suggest treating the energy fluctuations as the physical quantity ``detected" by gravity, when
one  incorporates quantum effects.  
If the \cc\ arises due to the energy density of the vacuum, then one needs to understand the structure of the quantum vacuum at cosmological scales. Quantum theory, especially the paradigm of renormalization group has taught us that the energy density --- and even the concept of the vacuum
state --- depends on the scale at which it is probed. The vacuum state which we use to study the
lattice vibrations in a solid, say, is not the same as vacuum state of the QED.

 In fact, it seems \textit{inevitable} that in a universe with two length scale $L_\Lambda,L_P$, the vacuum
 fluctuations will contribute an energy density of the correct order of magnitude $\rho_{_{\rm DE}}=\sqrt{\rho_{_{\rm IR}}\rho_{_{\rm UV}}}$. The hierarchy of energy scales in such a universe, as detected by 
 the gravitational field has \cite{plumian,tpvacfluc}
 the pattern
 \begin{equation}
\rho_{\rm vac}={\frac{1}{ L^4_P}}    
+{\frac{1}{L_P^4}\left(\frac{L_P}{L_\Lambda}\right)^2}  
+{\frac{1}{L_P^4}\left(\frac{L_P}{L_\Lambda}\right)^4}  
+  \cdots 
\label{rhoseries}
\end{equation}  
 The first term is the bulk energy density which needs to be renormalized away by an ad hoc process in the \textit{conventional} approaches. But in the approach outlined in the last section, we can ignore this because gravity is described by purely surface term in action. The third term is just the thermal energy density of the deSitter vacuum state; what is interesting is that quantum fluctuations in the matter fields \textit{inevitably generate the second term}.

The key new ingredient arises from the fact that the properties of the vacuum state  depends on the scale at which it is probed and it is not appropriate to ask questions without specifying this scale. 
 If the spacetime has a cosmological horizon which blocks information, the natural scale is provided by the size of the horizon,  $L_\Lambda$, and we should use observables defined within the accessible region. 
The operator $H(<L_\Lambda)$, corresponding to the total energy  inside
a region bounded by a cosmological horizon, will exhibit fluctuations  $\Delta E$ since vacuum state is not an eigenstate of 
{\it this} operator. The corresponding  fluctuations in the energy density, $\Delta\rho\propto (\Delta E)/L_\Lambda^3=f(L_P,L_\Lambda)$ will now depend on both the ultraviolet cutoff  $L_P$ as well as $L_\Lambda$.  
 To obtain
 $\Delta \rho_{\rm vac} \propto \Delta E/L_\Lambda^3$ which scales as $(L_P L_\Lambda)^{-2}$
 we need to have $(\Delta E)^2\propto L_P^{-4} L_\Lambda^2$; that is, the square of the energy fluctuations
 should scale as the surface area of the bounding surface which is provided by the  cosmic horizon.  
 Remarkably enough, a rigorous calculation \cite{tpvacfluc} of the dispersion in the energy shows that
 for $L_\Lambda \gg L_P$, the final result indeed has  the scaling 
 \begin{equation}
 (\Delta E )^2 = c_1 \frac{L_\Lambda^2}{L_P^4} 
 \label{deltae}
 \end{equation}
 where the constant $c_1$ depends on the manner in which ultra violet cutoff is imposed.
 Similar calculations have been done (with a completely different motivation, in the context of 
 entanglement entropy)
 by several people and it is known that the area scaling  found in Eq.~(\ref{deltae}), proportional to $
L_\Lambda^2$, is a generic feature \cite{area}.
For a simple exponential UV-cutoff, $c_1 = (1/30\pi^2)$ but  cannot be computed
 reliably without knowing the full theory.
  We thus find that the fluctuations in the energy density of the vacuum in a sphere of radius $L_\Lambda$ 
 is given by 
 \begin{equation}
 \Delta \rho_{\rm vac}  = \frac{\Delta E}{L_\Lambda^3} \propto L_P^{-2}L_\Lambda^{-2} \propto \frac{H_\Lambda^2}{G}
 \label{final}
 \end{equation}
 The numerical coefficient will depend on $c_1$ as well as the precise nature of infrared cutoff 
 radius (like whether it is $L_\Lambda$ or $L_\Lambda/2\pi$ etc.). It would be pretentious to cook up the factors
 to obtain the observed value for dark energy density. 
 But it is a fact of life that a fluctuation of magnitude $\Delta\rho_{vac}\simeq H_\Lambda^2/G$ will exist in the
energy density inside a sphere of radius $H_\Lambda^{-1}$ if Planck length is the UV cut off. {\it One cannot get away from it.}
On the other hand, observations suggest that there is a $\rho_{vac}$ of similar magnitude in the universe. It seems 
natural to identify the two, after subtracting out the mean value by hand. Our approach explains why there is a \textit{surviving} cosmological constant which satisfies 
$\rho_{_{\rm DE}}=\sqrt{\rho_{_{\rm IR}}\rho_{_{\rm UV}}}$
 which ---  in our opinion --- is {\it the} problem.

\section{ Conclusions}

It is obvious that  the existence of a component with negative pressure constitutes a major challenge in theoretical physics.
 The simplest choice for this component is the cosmological constant; other models based on scalar fields [as well as those based on branes etc. which I did not have time to discuss] do not alleviate the difficulties faced by \cc\  and --- in fact --- makes them worse. 
 The key point I want to stress is that the cosmological constant
  is most likely to be a low energy relic of a quantum gravitational effect or principle and its explanation will require a radical shift in our current paradigm.
  
  I have tried to advertise a new approach to gravity as a possible broad paradigm
  in which the observed value of the cosmological constant emerges naturally. The conceptual
  basis for this claim rests on the following ingredients.
  \begin{itemize}
\item The current problem of \cc\ is strongly dependent on how  we view the gravitational
degrees of freedom and its coupling to the vacuum energy. In the new  description
of gravity based only on the surface term, there are no bulk modes which couples to the vacuum and
the \textit{cosmological constant arises as an integration constant}. In such an approach, the 
bulk value of cosmological constant (the first term in \eq{rhoseries}) is irrelevant. 

\item The procedure works for a large class of theories (including Gauss-Bonnet type actions in higher dimensions) which are based on principle of equivalence and general covariance. This suggests that the mechanism for ignoring the bulk \cc\ is likely to survive quantum gravitational corrections which are likely to bring in additional, higher derivative, terms to the action.

\item Any generic null surface in a spacetime acts as a horizon for some class of observers.
Horizons modify the pattern of vacuum fluctuations  and macroscopic gravity acts as a detector of 
these fluctuations. When computed in a universe with asymptotically deSitter horizon, the 
vacuum fluctuations inside the horizon lead to the observed value of the cosmological constant (the second term in \eq{rhoseries}). 

\end{itemize} 

Getting the correct value of the cosmological constant (the second term in \eq{rhoseries})  is not as difficult as understanding why the bulk value (the first term in \eq{rhoseries} which is larger
by $10^{120}$!) can be ignored. It is possible to come up with different ad-hoc procedures to do this.
I want to emphasize that the holographic approach to gravity provides a natural backdrop for 
ignoring the bulk term --- and as a bonus --- we get the right value for the cosmological 
constant. It  is small because it is a purely quantum effect. 

This paradigm  treats continuum spacetime as analogous to a solid and Einstein's equations
as analogous to the elastic dynamics of the solid. This thermodynamic approach acquires surprising
support from the results I described in Section 5.  The fact that any theory based on principle of
equivalence and general covariance can be described by an action principle involving only the 
surface degrees of freedom cuts right into the heart of the matter. In all these theories of gravity,
\cc\ will emerge as an integration constant. Considering the generality of this approach, the physical
picture should be based on the ability of quantum micro-structure of spacetime to readjust itself
absorbing bulk vacuum energy density (like a sponge absorbing water). What one could observe
at macroscopic scales is the residual fluctuations (which is like the wetness of the sponge).

\section*{Acknowledgements}

I thank the organisers of the  Albert Einstein Century International Conference at Palais de l'Unesco, Paris, France, 18-23 July, 2005, especially Jean-Michel Alimi, for the warm hospitality during the conference.


\begin{thebibliography}{99}
\bibitem{note1} In all the expressions, while taking the derivatives, all components of $\Gamma^k_{cb}$ are treated as independent with no symmetry requirements. After the derivative is computed, one relates the 
$\Gamma^k_{cb}$s to the metric in the usual manner. Hence the order of lower indices of $\Gamma^k_{cb}$s in $L_{bulk}$ etc. is important.
\bibitem{clarify} More directly, note that any function $L$ can be written as 
$L=(L/2R)(\delta^c_ag^{bd}-\delta^d_ag^{bc})R^a_{\phantom{a}bcd}=Q_a^{\phantom{a}bcd}R^a_{\phantom{a}bcd}$ with $Q_a^{\phantom{a}bcd}=(L/2R)(\delta^c_ag^{bd}-\delta^d_ag^{bc})$.
\end{thebibliography}
\end{document}